\documentstyle[sprocl,epsfig]{article}

\def\ZPC{{\em Z. Phys.} C}

\def\be{\begin{equation}}
\def\ee{\end{equation}}
\def\bea{\begin{eqnarray}}
\def\eea{\end{eqnarray}}

\begin{document}
\begin{flushright}
Freiburg-THEP 99/18\\
LAPTH 745-99\\
August 1999 \vspace {0.8cm}\\
\end{flushright}

\title{The stealthy Higgs model at future Linear Colliders
\footnote{Presented at the Worldwide Study on Physics and Experiments
with Future Linear $e^+e^-$ Colliders, Sitges, Spain, april,may 1999.}}

\author{T. Binoth}

\address{Laboratoire d'Annecy--Le--Vieux de Physique Theorique LAPP,
Chemin de Bellevue, B.P. 110, F-74941, Annecy--Le--Vieux, France}

\author{ J.J. van der Bij}

\address{Fakult\"at f\"ur Physik, Albert--Ludwigs--Universit\"at Freiburg,
Hermann--Herder--Strasse 3, 79104 Freiburg, Germany}

\maketitle\abstracts{
We investigate the influence of scalar gauge singlets on Higgs signals
at a linear collider. These lead to a large invisible decay width of the Higgs.
We find that for high luminosities ($500-1000 fb^{-1}$) one can
essentially cover the allowed parameter range of the model.}

\section{Introduction}

Understanding of the electroweak symmetry breaking mechanism is one of the main
tasks in particle physics. The establishment of the structure of the Higgs sector would 
be a break-through in our knowledge about matter. So it is important to 
think about alternatives to the Standard Model Higgs sector especially
if they lead to a dilution of the signal.
The simplest possible
extension is the addition of scalar fields which are singlets under
the gauge group of the Standard Model. 
Radiative corrections to weak processes are not sensitive to the
presence of singlets in the theory, because no Feynman graphs containing
singlets  appear
at the one--loop level. Since effects at the two--loop level
are below the experimental precision,
the presence of a singlet sector is not ruled out by any 
of the LEP1 precision data. The only connection to such a hidden sector
is a possible Higgs singlet coupling, leading to a nonstandard invisible
Higgs decay. Whereas the invisible decay of the Higgs boson with a width comparable to
the Standard Model 
leads to relatively sharp missing energy signals, e.g. well known from discussions
on Majoron models \cite{valle}, a strongly coupled hidden sector could lead to fast
Higgs decay and thereby to wide resonances. This would disturb the signal to background
ratio if necessary cuts are imposed.  

To check the influence of a hidden sector we will study the coupling
of a Higgs boson to an O(N) symmetric set of scalars, which  
is one of  the simplest possibilities, introducing only a few extra 
parameters in the theory. The effect of the extra scalars is practically
the presence of a possibly large invisible decay width of the Higgs particle.
When the coupling is large enough the Higgs resonance can become
wide even for a light Higgs boson. It was shown earlier that there
is a range of parameters, where such a Higgs boson can be seen neither
at LEP nor at the LHC \cite{vladimir,lep2report,DPRoy}. 

\section{The model}
\begin{figure}[hbt]
\vspace{0.1cm}
\centerline{\epsfig{figure=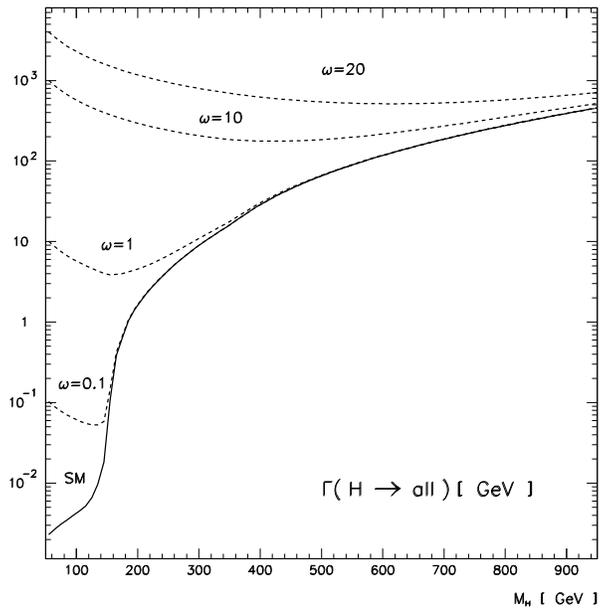,height=8.cm,angle=0}}
\caption{\it Higgs width in comparison with the Standard Model.}
\label{width}
\end{figure}

The scalar sector of the model consists of the usual Higgs sector coupled 
to a real N--component vector $\vec\varphi$ of scalar fields, denoted by 
Phions in the following. The lagrangian density is given by,
\bea
\label{definition}
 {\cal L}  &=&
 - \partial_{\mu}\phi^+ \partial^{\mu}\phi -\lambda (\phi^+\phi - v^2/2)^2
   - 1/2\,\partial_{\mu} \vec\varphi \partial^{\mu}\vec\varphi
     -1/2 \, m^2 \,\vec\varphi^2 \nonumber \\
     &&- \kappa/(8N) \, (\vec\varphi^2 )^2
    -\omega/(2\sqrt{N})\, \, \vec\varphi^2 \,\phi^+\phi \nonumber
\eea
where $\phi$ is the standard Higgs doublet. 
Couplings to fermions and vector bosons are the same as in the Standard Model.
The ordinary
Higgs field acquires the vacuum expectation value $v/\sqrt{2}$. For positive $\omega$
the $\vec\varphi$--field acquires no vacuum expectation
value. After spontaneous
symmetry breaking one is left with the ordinary Higgs boson,
coupled to the Phions into which it decays. Also the Phions
receive an induced mass from the spontaneous symmetry breaking which is suppressed
by a factor $1/\sqrt{N}$.
If the factor N is taken
to be large,  the model can be analysed with $1/N$--expansion techniques.
By taking this limit the Phion mass is suppressed, whereas
the decay width of the Higgs boson is not. Because the Higgs width
is now depending on the Higgs Phion coupling its value is arbitrary. 
Therefore the main effect of the presence of the Phions is to give
a possibly large invisible decay rate to the Higgs boson. The 
invisible decay width is given by 
\be \Gamma_H =\frac {\omega^2 v^2}{32 \pi M_H} = 
\frac {\omega^2 (\sin\theta_W\cos\theta_W M_Z)^2)}{32 \pi^2 \alpha_{em} M_H}\quad .\nonumber \ee
The Higgs width is compared with the width in the Standard Model for various choices
of the coupling $\omega$ in Fig.~\ref{width}.
The model is different
from Majoron models \cite{valle}, since the width is not necessarily small.
The model is similar to the technicolor--like model of Ref.~\cite{chivukula}.

\begin{figure}[htb]
\vspace{0.1cm}
\centerline{\epsfig{figure=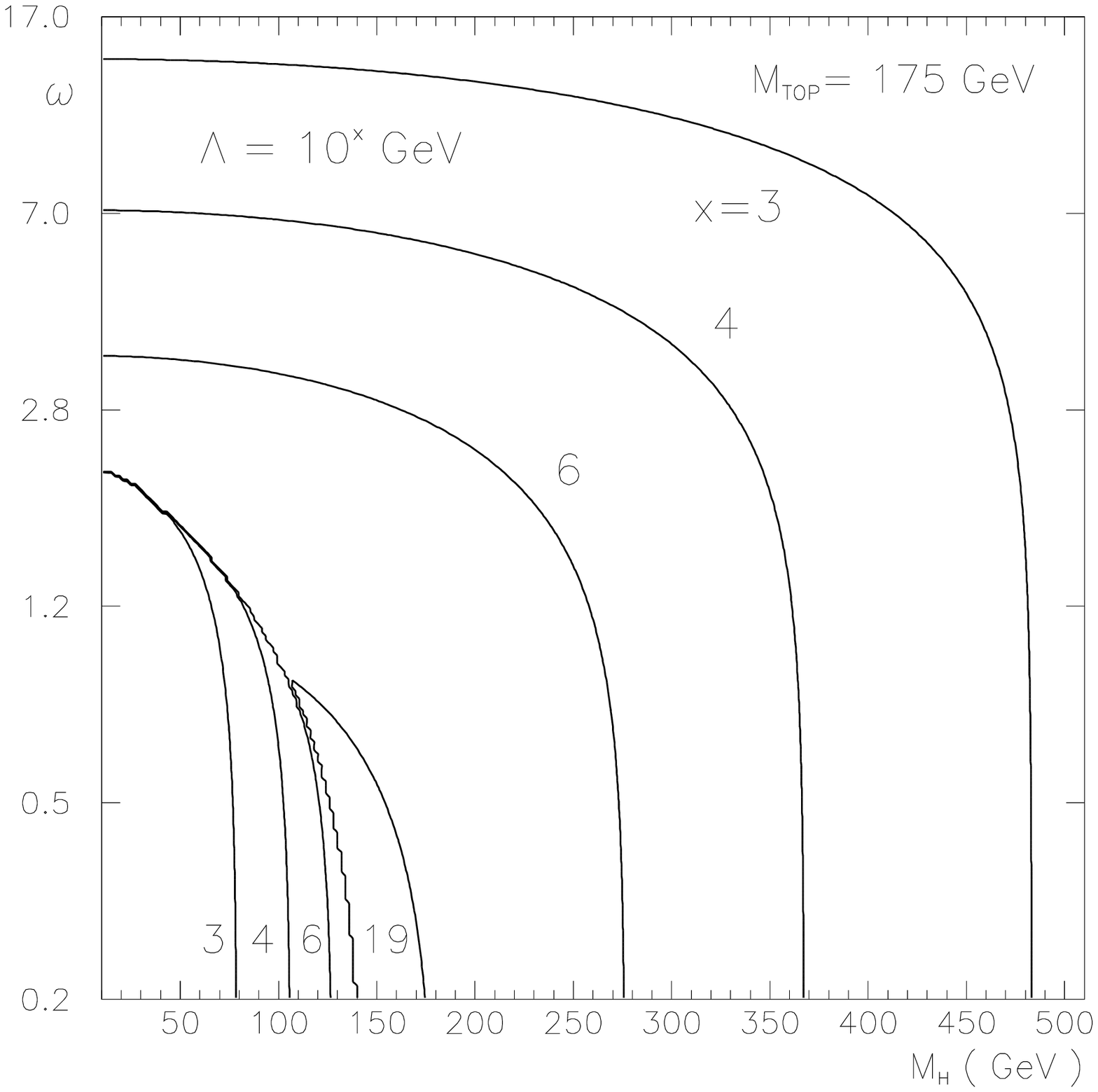,height=8.cm,angle=0}}
\caption{\it Theoretical limits on the parameters of the model
in the $\omega$ vs. $M_H$ plane. The contour lines correspond 
to the cutoff scales $\Lambda = 10^{19}$, $10^6$, $10^4$ and $10^3$ GeV.}
\label{stability}
\end{figure}

Consistency of the model requires two conditions.
One condition is the absence of a Landau pole below a certain scale
$\Lambda$. The other follows from the stability of the vacuum up to a certain
scale. An example of such limits is given in Fig.~\ref{stability},
where $\kappa=0$ was taken at the scale $2m_Z$, which allows for
the widest parameter range.
The regions of validity up to a given scale $\Lambda$ is sandwiched
between the upper--right and the lower--left contour lines in the figure. 
The first stem from the Landau pole, the second from instability of the 
vacuum at that scale.

To search for the Higgs boson there are basically
two channels, one is the standard decay, which is reduced in branching
ratio due to the decay into Phions.
The other is the invisible decay, which rapidly becomes dominant,
eventually making the Higgs resonance wide (see Fig.~\ref{width}). 
In order to give the bounds we 
neglect the coupling $\kappa$ as this is a small effect. We
also neglect the Phion mass. (For other values of the Phion mass
the bounds can be found by rescaling the decay widths
with the appropriate phase space factor.) 

\section{LC bounds}
At a linear collider (LC) the upper limits on the couplings in the present model
come essentially from the invisible decay, as the branching ratio
into visible particles drops with increasing $\varphi$--Higgs
coupling, whereas for the Higgs mass limits one has to consider visible decays, too.
The $WW$--fusion process
can not be used to look for invisible Higgs decay.
One is therefore left with the Higgsstrahlung und $ZZ$--fusion reaction. 
For energies up to 500 GeV the Higgsstrahlungs cross section is dominant and
still comparable if one multiplies with the branching ratio $B(Z\rightarrow e^+e^-,\mu^+\mu^-)$. 
The Higgsstrahlungs reaction is preferred, because one can tag the on-shell Z boson.
 Thus we only have considered reactions 
containing an on shell Z boson with its decay into $e^+e^-$ or $\mu^+\mu^-$.  
The signal cross section is the well  known Higgsstrahlungs cross section modified
by the non standard Higgs width due to Phion decay. With the invariant mass of the invisible
Phion system, $s_I$, it reads:  

\be \sigma_{(e^+e^-\rightarrow Z+E\!\!\!/)} = 
\int ds_I \, \sigma_{(e^+e^-\rightarrow ZH)}(s_I) \,
\frac{\sqrt{s_I} \quad \Gamma(H\rightarrow E\!\!\!/)}
{\pi ((M_H^2-s_I)^2+s_I\,\Gamma(H\rightarrow \mbox{All})^2)}\nonumber\ee

To reduce the $Z \nu\nu$ background \cite{mele}, we used the fact
that the angular distribution of the Z--boson for the signal peaks for small values
of $|\cos\theta_Z|$ in contrast to the background. Thus we imposed the cut $|\cos\theta_Z|<0.7$.
Because we assume the reconstruction of the on-shell Z--boson we use the kinematical relation
\newline $E_Z=(\sqrt{s}-M_Z^2+s_I)/(2\sqrt{s})$
between the Z energy and the invariant mass of the invisible system
to define a second cut. Because the differential cross section $d\sigma/ds_I$ peaks
at $M_H^2$, we impose the following condition on the Z energy:
\be \frac{\sqrt{s}-M_Z^2+(M_H-\Delta_H)^2}{2\sqrt{s}}<E_Z<
\frac{\sqrt{s}-M_Z^2+(M_H-\Delta_H)^2}{2\sqrt{s}}\nonumber\ee 
For the choice of $\Delta_H$ a comment is in order. As long as the Higgs width is small one
is allowed to use small  $\Delta_H$, which reduces the background considerably keeping
most of the signal events. But in the case of large $\varphi$--Higgs coupling, $\omega$, one
looses valuable events. To compromise between both effects we took  
$\Delta_H=30 (100)$ GeV for colliders with center of mass energy of
$500 (1400)$ GeV, respectively.   
 
\begin{figure}[htb]
\vspace{0.1cm}
\centerline{\epsfig{figure=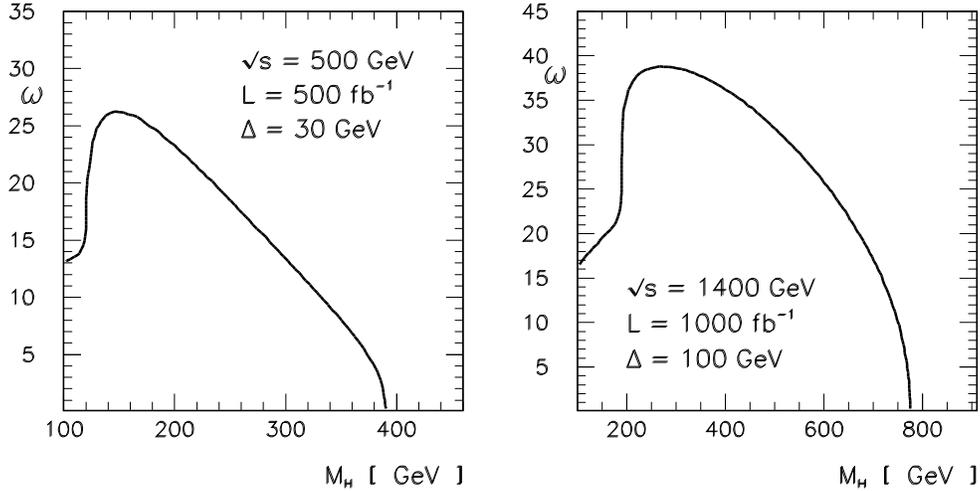,height=6.5cm,angle=0}}
\caption{\it Exclusion limits at a LC at an energy of 500 (1400) GeV and
luminosity 500 (1000) $fb^{-1}$, respectively. }
\label{exclu1}
\end{figure}

For the exclusion limits we assumed an integrated luminosity
of $500$ ($1000$) $fb^{-1}$ for the two center of mass energies. 
To define the $95 \%$ confidence level we used 
Poisson statistics as in Ref. \cite{lep2report}.
The result is given  
in Fig.~\ref{exclu1}.

We conclude from the above that a LC with the proposed 
high luminosities can
essentially cover the parameter range up to the theoretically allowed 
limit with a
completely clean signal, consisting of leptons plus missing energy. 
Such a LC appears to be the 
unique machine to be sensitive to this class of models.
\newpage

\section*{Acknowledgements}
This work was supported by the DFG-Forschergruppe Quantenfeldtheorie,
Computeralgebra und Monte Carlo Simulation, the EU grant 
FMRX-CT98-0194(DG12-MIHT) and by the NATO-grant 
CRG 970113.

\section*{References}
\begin{thebibliography}{99}
\bibitem{vladimir}
T.~Binoth, J.~J.~van~der~Bij, \ZPC75, 17 (1997) and references therein.
\bibitem{valle}
J.~Valle et al. LEP2 Higgs Report, CERN 96-01, 350 (1996).
\bibitem{DPRoy}
D.~Choudhury, D.~P.~Roy, Phys.Lett. B322, 368 (1994).
\bibitem{eboli}
O.~J.~P.~Eboli et al., DESY 93-123C, 55 (1993). 
\bibitem{chivukula}
R.~S.~Chivukula, M.~Golden, Phys. Lett. B267, 233 (1991);\\
J.~D.~Bjorken, Int. J. Mod. Phys. A7, 4221 (1992).
\bibitem{lep2report} LEP2 Higgs Report, CERN 96-01, 350 (1996).
\bibitem{mele} Ambrosanio S., Mele B., Nucl. Phys. B374, 3 (1992).
\end{thebibliography}
\end{document}